\documentclass[twocolumn]{revtex4-2}
\usepackage[dvipdfmx]{graphicx}
\usepackage{amsmath,amsbsy,amssymb}
\usepackage{bm}
\usepackage{mathrsfs}
\usepackage{ulem}
\usepackage{textgreek}
\usepackage{mathrsfs}
\usepackage{braket}
\usepackage{physics}

\usepackage{color}

\newcommand{\diff}{\mathrm{d}}

\newcommand{\imu}{\mathrm{i}}
\newcommand{\epn}{\mathrm{e}}

\newcommand{\ua}{\uparrow}
\newcommand{\da}{\downarrow}
\newcommand{\dg}{\dagger}
\newcommand{\la}{\langle}
\newcommand{\ra}{\rangle}
\newcommand{\al}{\alpha}
\newcommand{\sg}{\sigma}
\newcommand{\gm}{\gamma}
\newcommand{\ep}{\varepsilon}

\newcommand{\comb}[2]{{}_{#1}\textrm{C}_{#2}}

\newcommand{\lam}{\lambda}

\begin{document}

\title{
 Material-based analysis of spin-orbital Mott insulators
}

\author{
Ryuta Iwazaki, 
Hiroshi Shinaoka, and
Shintaro Hoshino
}

\affiliation{
Department of Physics, Saitama University, Shimo-Okubo, Saitama 338-8570, Japan
}

\date{\today}

\begin{abstract}

We present a framework for analyzing Mott insulators using a material-based tight-binding model.
We start with a realistic multiorbital Hubbard model and derive an effective model for the localized electrons through the second-order perturbation theory with respect to intersite hopping.
This effective model, known as the Kugel-Khomskii model, is described by SU($N$) generators, where $N$ is the number of localized states.  
We solve this model by the mean-field theory that takes local correlations into account and reveal spin-orbital ordered states.
To include spatial correlations, we apply the classical Monte Carlo based on the path-integral approach with SU($N$) coherent states, and also derive the equation of motion for spin-orbital degrees of freedom.
Our approach is applicable to any Mott insulator with reasonable computational cost.
The $5d$-pyrochlore oxide is used here as demonstration.

\end{abstract}

\maketitle

\textit{Introduction}.---
Multiorbital systems with strongly correlated electrons have been attracting attention due to their diverse physical phenomena, such as electronic ordering and multiferroic behavior.
It is crucial to uncover their material-specific physical properties in order to make a serious comparison with experimental results.
In materials with weakly correlated electrons, density functional theory (DFT)-based calculations have been successful in describing their electronic properties.
On the other hand, in the strongly correlated regime, it is useful to construct a tight-binding model using localized Wannier functions and subsequently employ a multiorbital Hubbard model with local Coulomb repulsive interactions as a fundamental model.
Unfortunately, it is extremely difficult to perform the calculations in a realistic setting due to the immense numerical cost.
A theoretical framework that is applicable to realistic strongly correlated electron systems is highly desired, which will enable material prediction through, for example, high-throughput screening~\cite{Jain2016}.

In the present work, we focus on the Mott insulators where the electrons are localized with strong local Coulomb interaction.
Even in this case, the spin-orbital degrees of freedom generate a number of interesting phenomena such as magnetic orderings, multiferroic behaviors and spin liquids~\cite{Imada1998, Tokura2000, Kim2008, Jackeli2009, Kim2009, Krempa2014, Kitagawa2018, Kasahara2018, Tang2022}.
The low-energy effective model with localized electrons is known as the Kugel-Khomskii model, in which both the spin and orbital degrees of freedom are involved~\cite{Kugel1972, Kugel1973, Cyrot1975, Khaliullin1997, Ishihara1997, Feiner1999, Ishihara2000, Harris2004, Ishihara2004, Normand2008, Nasu2013, Koga2018, Otsuki2019, Bieniasz2019, Nasu2021, Khaliullin2021, Khomskii2022}.
The realistic localized models have been discussed for the spin model~\cite{Zhang2012, Chiesa2013, Yamaji2014prl, Rau2014, Winter2016, Winter2017, Kurzydwski2017, Chiesa2019, Huang2020, Kaib2021, Churchill2022, Mosca2022} and 
$e_g$/$t_{2g}$-multiorbital systems~\cite{Pavarini2008, Pavarini2010, Autieri2014, Snamina2016, Jeanneau2017, Aligia2019, Zhang2022}.
The DFT+DMFT approaches have also been employed for the analysis \cite{Pavarini2008, Pavarini2010, Zhang2012, Autieri2014, Pourovskii2021, Mosca2021, Mosca2022, Pourovskii2022arxiv}.
In order to study {\it arbitrary} Mott insulator materials, however, a more general framework is needed that can be applied at reasonable computational cost to general multiorbital systems with spin-orbit interactions and any number $N$ of localized states per atom.

In this paper, we propose a general framework to perform calculations for the spin-orbital Mott insulators, which is not restricted to specific systems.
We develop a realistic Kugel-Khomskii model based on the tight-binding model derived from the first-principles calculation and the local Coulomb interaction with Slater-Condon parameters.
The model contains $N^2-1$ spin-orbital degrees of freedom and is described by SU($N$) generators.
When analyzing the model, while a fully quantum analysis is not feasible because of a huge computational cost, we use the classical Monte Carlo with the SU($N$) coherent state~\cite{Perelomov1972, Gnutzmann1998, Nemoto2000}, in addition to the standard mean-field theory.
The SU($N$) coherent state has been used for the spin systems~\cite{Read1989, Stoudenmire2009, Zhang2021, Remund2022, Dahlbom2022Aug, Seifert2022, Dahlbom2022Dec, HZhang2022arxiv, Do2022arxiv, Pohle2022arxiv}, and here we apply it to the realistic Kugel-Khomskii model.
While the quantum mechanical inter-site correlations at very low temperatures are not incorporated in our theory, our method captures the characteristic physics at finite temperatures in a realistic setup for any Mott insulators with reasonable numerical cost.

\begin{figure}
    \centering
    \includegraphics[width = 85mm]{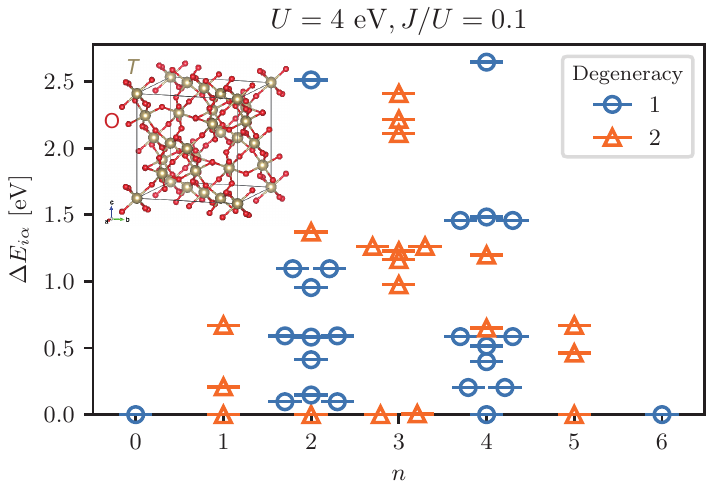}
    \caption{
        Single-site eigenenergy levels for Cd$_2$$T_2$O$_7$, which corresponds to $\mathscr H_t = 0$.
        We choose $U = 4$~eV and $J/U = 0.1$ which is comparable to the previous study~\cite{Shinaoka2015}.
        The vertical axis shows the energy measured from the lowest energy at each $n$.
        The inset is the crystal structure of Cd$_2T_2$O$_7$, where only $T$ and O atoms are shown for clarity~\cite{Momma2011}.
    }
    \label{fig:ene_level}
\end{figure}

We will take the pyrochlore oxide Cd$_2T_2$O$_7$ as an example.
This is suitable as a prototype material for the demonstration of our framework due to its complicated electronic structure: the four transition metal $T$ atoms in unit cell (specified as sublattice indices A, B, C, D), large spin-orbit interaction, and $t_{2g}$ three orbitals of $5d$ electrons with trigonal symmetry at $T$ atom site~\cite{Gardner2010, Shinaoka2019} (see Fig.~\ref{fig:ene_level}).
In addition, their non-colinear magnetic structures are well studied both theoretically and experimentally~\cite{Shinaoka2012, Yamaura2012}.
Hence the applicability to this prototypical material Cd$_2T_2$O$_7$ demonstrates the versatility of our method.

\textit{Realistic Kugel-Khomski model}.---
The realistic effective model for the localized electrons are constructed based on the multiorbital Hubbard model derived from the first principles calculation.
Let us begin with the Hamiltonian $\mathscr{H} = \mathscr{H}_{\mathrm{loc}} + \mathscr{H}_t$, where
\begin{align}
\mathscr{H}_t = \sum_{\la ij\ra} \sum_{ab} t^{ab}_{ij} c_{ia}^\dg c_{jb} + \mathrm{H.c.}
\end{align}
describes the intersite hopping term.
The operator $c_{ia}$ annihilates the electron at the atom site $i$ with the spin($\sg$)-orbital($\gm$) index $a=(\gm,\sg)$.
The symbol $\la ij\ra$ indicates the summation with respect to the pairs of atomic sites, and includes the terms other than the nearest neighbor sites.
The local part $\mathscr{H}_{\mathrm{loc}}$ is further divided into three components as $\mathscr{H}_{\mathrm{loc}} = \mathscr{H}_U + \mathscr{H}_{\mathrm{SOC}} + \mathscr{H}_{\mathrm{CEF}}$, which are the Coulomb interaction, the spin-orbit coupling and the local crystalline electric field, respectively.
The Coulomb interaction is written as
\begin{align}
    \mathscr{H}_U
    =
    \sum_{i \gm_1 \gm_2 \gm_3 \gm_4 \sg \sg'} U_{\gm_1 \gm_2 \gm_3 \gm_4} c_{i\gm_1 \sg}^\dg c_{i\gm_2 \sg'}^\dg c_{i\gm_4 \sg'} c_{i\gm_3\sg},
\end{align}
which is parameterized by the Slater-Condon parameters as typically used in LDA+$U$ or LDA+DMFT framework~\cite{Kotliar2006}.
Specifically for the three orbital case as in $t_{2g}$ orbital, the standard Slater-Kanamori form is employed:
$U_{\gm\gm\gm\gm} = U/2$, $U_{\gm\gm'\gm\gm'} = U'/2$, $U_{\gm\gm'\gm'\gm} = U_{\gm\gm\gm'\gm'} = J/2$ for $\gm \neq \gm'$ ($U'=U-2J$) and the other terms are zero.

\begin{figure}
    \centering
    \includegraphics[width = 85mm]{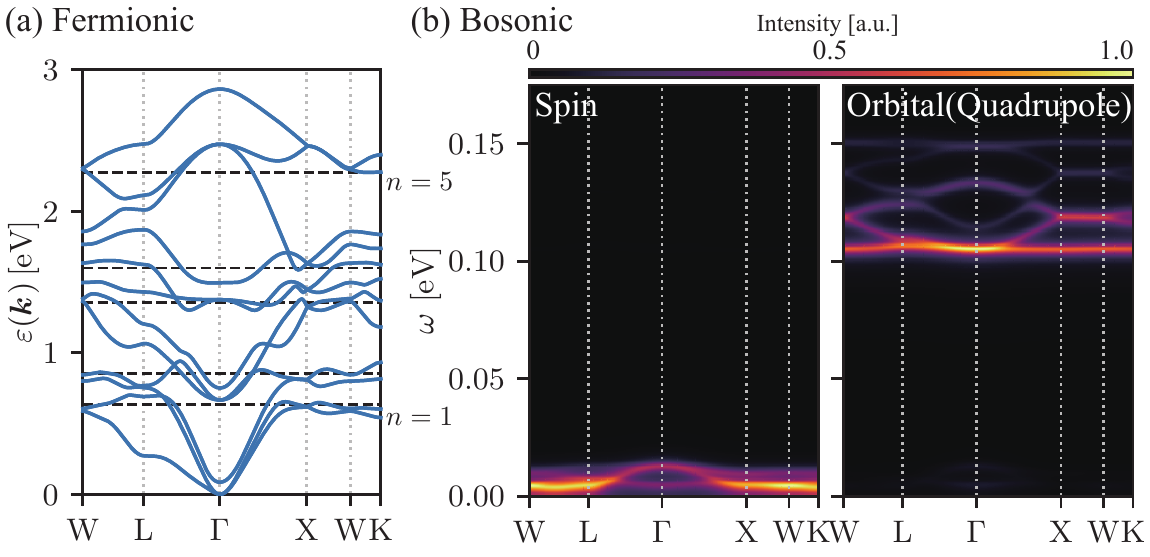}
    \caption{
        (a) Electronic energy band for Cd$_2T_2$O$_7$.
        The vertical axis is measured from the bottom of the bands.
        The horizontal dashed lines express the chemical potential for each $n$.
        (b) Bosonic energy spectra $\Im \chi (\bm{q}, \omega)/\omega$ for the SU(6) model at $T = 10^{-3}$~eV.
        The left panel shows the dispersion for the spin, while the orbital excitation spectra is shown in the right panel.
    }
    \label{fig:n1_mf_qdep}
\end{figure}

In the following, we take the tight-binding model of $T=$ Os derived from electronic-structure calculation~\cite{supple}.
Since the band structure [see Fig.~\ref{fig:n1_mf_qdep}(a)] is similar to the other materials with different filling such as $T=$ Re~\cite{Singh2002, Harima2002}, we use the data of the $T=$ Os case also for the other electron fillings.

We analyze the multiorbital Hubbard model in the strong coupling limit ($U\to \infty$), where the electrons are localized.
First of all, we derive the eigenenergies and eigenfunctions in the atomic model with only $\mathscr H_{\rm loc}$, which is necessary for specifying the model Hilbert space at low energies, \textit{i.e.}, the number $N$ of the localized states.
Figure~\ref{fig:ene_level} shows the single-site eigenenergy diagram of $\mathscr{H}_{\mathrm{loc}}$ for each number $n$ of electrons per $T$ atom.
When we focus on the odd number of the filling $n$, there are only doubly degenerated states corresponding to the Kramers doublet.

In this paper, we choose $n=1$ for a demonstration of our scheme, which allows us to choose the size of the model space as $N=2$, $4$, $6$ based on Fig.~\ref{fig:ene_level}.
We call them SU(2), SU(4) and SU(6) models, respectively.
The SU($N$) model contains $N^2-1$ operators for each atom.
The procedure for the simplest $N=2$ case is summarized in Supplementary Matreial (SM)~\cite{supple}.
Although the dimension of the model Hilbert space may be dependent on the lattice site, we here take the same $N$ for all the sites.

Once the model space is specified, we treat the intersite Hamiltonian $\mathscr{H}_t$ as a perturbation, to obtain the effective Hamiltonian which gives a correct eigenenergies within the restricted Hilbert space~\cite{Okubo1954, Bloch1958, desCloizeaux1960, Durand1983, KuramotoBook2020}.
While there are several choices of the form of the effective Hamiltonian, the Hermitian Hamiltonian (des Cloizeaux type) is easier to be handled~\cite{desCloizeaux1960, Durand1983}.
We focus on the two atoms which are connected by the hopping matrix $\mathscr H_t$,
and expand this two-site Hamiltonian up to second order of $\mathscr H_t$~\cite{supple}.
Thereby we obtain the matrix element of the effective Hamiltonian whose size is $N^2\times N^2$.
We can rewrite the obtained effective Hamiltonian by complete local operators $\mathscr O_i$ at the site $i$.
We employ the numerical calculation with matrix multiplications for this procedure~\cite{Iwazaki2021}.
Collecting all the combinations of the two-site Hamiltonians, we obtain the following realistic Kugel-Khomskii model: 
\begin{align} \label{eq:localized_model}
    \mathscr{H}_{\mathrm{eff}}[\mathscr O]
    =
    \sum_{\la ij \ra}\sum_{\xi\xi'} I_{ij}^{\xi\xi'} \mathscr{O}_i^\xi \mathscr{O}_j^{\xi'}
    -
    \sum_i \sum_\xi H_i^\xi \mathscr{O}_i^\xi
    ,
\end{align}
where both the zeroth- and second-order contributions are involved in this effective Hamiltonian.
We have defined the local operators $\mathscr{O}_i^\xi = \sum_{\al\beta} O_{\al\beta}^\xi |\al\ra_i \, {}_i\la\beta|$ ($\al=1,\cdots, N$, $\xi = 0, \cdots, N^2-1$), where $|\al\ra_i$ is a state vector in the model Hilbert space at site $i$.
We use the matrices $O^\xi_{\al\beta}$ with completeness and orthonormality (\textit{e.g.} for single orbital model, we take the SU(2) generators, which are the Pauli matrices)~\cite{supple}.
We emphasize that this Hamiltonian is derived from the first-principles calculation data, where the tunable parameters are only the local Coulomb interaction parameters $U$ and $J$.
In the actual calculation, the data of $I_{ij}^{\xi\xi'}$ is outputted with the data structure similar to the original input of $t_{ij}^{ab}$.

Since it is in general difficult to interpret the physical meaning of the local operators $\mathscr{O}_i^\xi$, it is desirable to transform them into physical quantities defined in terms of the original electronic system.
Let us consider the local physical quantity $\mathscr A_i$.
This can be spin or orbital operator if we choose the form of $\mathscr{A}_i = \frac 1 2 \sum_{ab} A_{ab} c_{ia}^\dg c_{ib}$ where the matrix $A$ is composed of a direct product of the matrices in spin and orbital spaces.
By using the projection operator onto the model Hilbert space, $\mathscr{P} = \prod_i \sum_{\al} |\al\ra_i\, {}_i\la\al|$, we obtain 
\begin{align} \label{eq:calc_qty}
    \mathscr P \mathscr A_i \mathscr P
    &= \sum_{\xi} \mathscr O_i^\xi \sum_{\al\beta} {}_i\la\al| \mathscr A_i |\beta\ra_i O^\xi_{\beta \al}
    .
\end{align}
We can get the matrix element ${}_i\la\al| \mathscr A_i |\beta\ra_i$ by analyzing $\mathscr H_{\rm loc}$.
Thus, once the expectation value of $\mathscr O_i$ is obtained by solving the model in Eq.~\eqref{eq:localized_model}, any local physical quantities can be evaluated through this formula.
It is notable that $\mathscr{A}_i$ can be chosen as many-body quantities such as a double occupancy, which is not usually considered for the conventional Kugel-Khomskii model.

The correlation functions are also useful quantities.
When we consider the linear response against a small fictitious field conjugate to $\mathscr{O}_i^\xi$, the dynamical susceptibilities are given by
\begin{align}
    \chi_{ij}^{\xi\xi'} (\imu\nu)
    =
    \int_0^{1/T} \diff\tau\, \qty[
        \la \mathscr{O}_i^\xi(\tau) \mathscr{O}_j^{\xi'}\ra - \la \mathscr{O}_i^\xi \ra \la \mathscr{O}_j^{\xi'}\ra
    ] \epn^{\imu\nu\tau},
    \label{eq:chidef}
\end{align}
where $\mathscr{O}_i^\xi(\tau) = \epn^{\tau\mathscr{H}} \mathscr{O}_i^\xi \epn^{-\tau\mathscr{H}}$, and $\tau$ is a Heisenberg picture with imaginary time, and $\nu=2\pi m T$ ($m\in \mathbb Z$) is a bosonic Matsubara frequency.
We have taken $k_\mathrm{B}=1$.
Using Eq.~\eqref{eq:calc_qty}, the susceptibility can be transformed into the physical susceptibilities defined in terms of the original electron operators.
The information of any spin-orbital excitation is encoded in Eq.~\eqref{eq:chidef}.
For example, we can obtain the dispersion of the orbiton, which is a quasiparticle describing the excitation of the orbital~\cite{Cyrot1975, Ishihara2000, Ishihara2004}.

\textit{Mean-field theory}.--- 
Since the obtained localized model contains quantum effects, it is still very hard to be solved.
In the following, we introduce several approximated methods to solve the realistic Kugel-Khomskii model given in Eq.~\eqref{eq:localized_model}.
The most fundamental approximation is the mean-field theory.
Defining the effective field $\tilde{H}_i^\xi = H_i^\xi - \sum_{j \neq i, \xi'} I_{ij}^{\xi\xi'} M_{j}^{\xi'}$, the mean-field Hamiltonian is written as 
\begin{align}
    \mathscr{H}_{\mathrm{MF}}
    =
    -\sum_i \sum_\xi \tilde{H}_i^\xi \mathscr{O}_i^\xi
    -\sum_{\la ij \ra} \sum_{\xi\xi'} I_{ij}^{\xi\xi'} M_i^\xi M_j^{\xi'}.
\end{align}
We have defined $M_i^\xi = \la \mathscr{O}_i^\xi \ra_{\mathrm{MF}}$ where the expectation value is taken as $\la \cdots \ra_{\mathrm{MF}} = \Tr (\cdots \epn^{-\mathscr{H}_{\mathrm{MF}}/T}) / \Tr \epn^{-\mathscr{H}_{\mathrm{MF}}/T}$.
We also evaluate the dynamical susceptibilities with the random phase approximation as
\begin{align}
    \hat{\chi}(\bm{q}, \omega)
    =
    \hat{\chi}_0(\omega) \qty[
        \hat{1} + \hat{I}(\bm{q}) \hat{\chi}_0(\omega)
    ]^{-1},
\end{align}
where the hat ($\,\hat{}\,$) symbol represents the matrix with respect to the index $\xi$, and $\hat{1}$ is the identity matrix.
We have defined the local susceptibility by $\hat{\chi}_{0}(\omega) = \hat{\chi}_{ii}(\omega + \imu 0^+) $ which is evaluated by the local mean-field Hamiltonian.

First of all, we show in Fig.~\ref{fig:n1_mf_qdep}(b) the spin-orbital excitation spectra of the realistic Kugel-Khomskii model, which is contrasted against the fermionic excitation of the original tight-binding electrons in (a).
We take the SU(6) model at $n = 1$ and $T = 10^{-3}$~eV.
The left panel of (b) is the spectra for the spin, which corresponds to the dispersion of the magnon.
The gapped excitation reflects the presence of the spin-orbit coupling.
The right panel is the spectra for the non-magnetic orbital (quadrupole) moment (see Ref.~\cite{supple} for the definition of the orbital moment).
This orbital excitation is unique to the SU(6) model, although the magnon dispersion is captured already in the SU(2) model.

\begin{figure}
    \centering
    \includegraphics[width = 85mm]{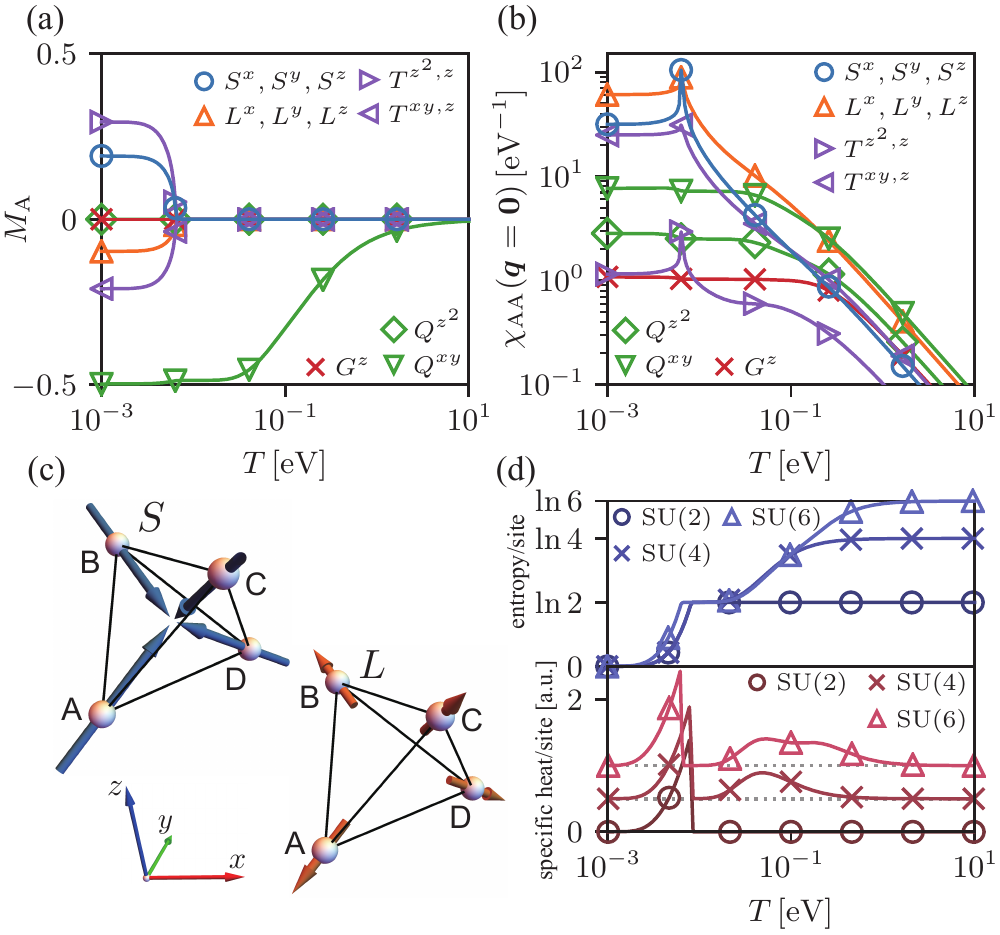}
    \caption{
        Temperature dependence of (a) the order parameters and (b) the $\bm{q} = \bm{0}$ component of the static diagonal susceptibilities at A-site for the SU(6) model obtained by the mean-field analysis.
        The blue, orange, green, red and purple lines show the spin, magnetic orbital, electric orbital, electric dipole and magnetic octupole, respectively.
        (c) Sketches for the spin ($S$) and the magnetic orbital moment ($L$) of the SU(6) model at $T = 10^{-3}$~eV.
        (d) Temperature dependence of the entropy (upper panel) and the specific heat (lower panel) with the circles, crosses and triangles for the SU(2), SU(4) and SU(6) models, respectively.
        For clarity, specific heat is vertically shifted for the SU(4,6) models.
    }
    \label{fig:n1_mf_tdep}
\end{figure}

We show the temperature dependence of the order parameters at A-sublattice in Fig.~\ref{fig:n1_mf_tdep}(a) for the SU(6) model.
The symbols $S, L, Q, G$ and $T$ are the spin, magnetic orbital, electric orbital (quadrupole), electric dipole and magnetic octupole moments, respectively~\cite{supple}.
At low temperatures with $T \lesssim 10^{-2}$~eV, the magnetic ordering occurs, whose order parameters are described by $S, L$ and $T$.
We also show the $\bm{q} = \bm{0}$ component of the diagonal susceptibilities at A-sublattice in Fig.~\ref{fig:n1_mf_tdep}(b), where the magnetic susceptibilities ($S$, $L$, $T$) diverge.
The magnetic structures at $T = 10^{-3}$~eV are shown at Fig.~\ref{fig:n1_mf_tdep}(c), which displays the all-in-all-out (AIAO) structure and the antiparallel alignment of $S$ and $L$ moment.
The AIAO-type magnetic ordering in $5d$ pyrochlore oxides have been suggested both theoretically and experimentally~\cite{Tomiyasu2012, Shinaoka2012, Yamaura2012, Sagayama2013, Disseler2014, Shinaoka2015}.

The temperature dependence of the thermodynamic quantities per site are shown in Fig.~\ref{fig:n1_mf_tdep}(d), with which we compare the results of the SU(2,4,6) models.
All of the models have an anomaly in the specific heat (lower panel) around $T_{\mathrm{c}} \simeq 10^{-2}$~eV, which signals a second-order phase transition.
The SU(6) model has a smaller magnetic transition temperature compared to the SU(2,4) cases.
The single site entropy (upper panel) has a $\ln 2$ plateau for the SU(4) and the SU(6) model just above $T_{\mathrm{c}}$, and it deviates from $\ln 2$ reflecting the additional degrees of freedom at higher $T$.
The specific heat above $T_{\mathrm{c}}$ shows Schottky peaks originating from the local energy-level splitting.

\textit{Classical model}.---We can also solve the model by applying the classical approximation to Eq.~\eqref{eq:localized_model}.
In this method, we can examine the effect of the non-local correlation. 
We employ the path-integral formalism using a coherent state~\cite{Nemoto2000, Zhang2021}, with which we derive both the classical partition function and equations of motion.
The coherent state is defined for each site $i$ by 
\begin{align}
|\Omega_i \ra
= \sum_{\al=1}^N c_\al(\Omega_i) |\al\ra_i,
\label{eq:coherent_state_def}
\end{align}
where $|\al\ra_i$ is a quantum state basis.
$\Omega_i$ is a set of local continuous variables:
$\Omega_i = \{ \xi_{1i}, \cdots, \xi_{N-1,i}, \varphi_{1i}, \cdots, \varphi_{N-1,i} \} $, each of which is written as $\Omega_{pi}$ ($p=1,\cdots,2(N-1)$)~\cite{supple}.
Here $\xi_{1i,\cdots} \in [0,\pi/2]$ and $\varphi_{1i,\cdots}\in [0,2\pi)$ respectively correspond to the generalized versions of polar angle and azimuthal angle of the spin in the SU(2) model.

The partition function is written as $Z = \int \mathscr D \bm \Omega\, \epn^{-\mathscr{S}}$, where the action is~\cite{Read1989}
\begin{align}
    \mathscr{S} &= \int \diff \tau\, \qty(
    \la \bm \Omega |\partial_\tau |\bm \Omega \ra +\la \bm \Omega| \mathscr H_{\rm eff} |\bm \Omega\ra
    ).
    \label{eq:action}
\end{align}
We have defined $|\bm \Omega\ra = \prod_{i}|\Omega_i\ra$ at an imaginary time $\tau$.
The quantum-mechanical operator $\mathscr O_i$ is now replaced by the classical variable:
$\mathcal O^\xi (\Omega_i) = \la \bm \Omega| \mathscr O_i^\xi |\bm \Omega \ra$.
Based on these expressions, the classical model can be rigorously derived by using the coherent state path integral method that omits the Berry phase term, as in the spin model~\cite{AuerbachBook1994}.
We can also show that the classical free energy is always larger than the quantum one~\cite{Lieb1973}, and it is ensured that the lowest-free-energy state in the classical model is energetically closest to the genuine quantum state.

The model can be numerically simulated by using the classical Monte Carlo method. 
We use the local Metropolis update and the replica exchange method which allow us to simulate the systems with various temperatures efficiently~\cite{Hukushima1996}.
In addition, we also apply the over-relaxation update~\cite{Creutz1987} for the more efficient simulation.
The over-relaxation update in the present case consists of microcanonical moves that does not alter the energy.
For the SU(2) case, the local spin vector is rotated around the local effective field by the angle $\pi$~\cite{LandauBook2021, Alonso1996}.
However, this cannot be directly extended to SU($N$) case, and the consideration based on the coherent state is needed.

To perform the over-relaxation update for the SU($N$) case, let us focus on the one lattice site $i$, and then its effective local Hamiltonian is written as
$    \mathcal H_{{\rm loc},i}  = - \sum_\xi \tilde H_i^\xi \mathcal O^\xi(\Omega_i)
$
where the effect of the surrounding sites is included in
$\tilde H_i^\xi = H_i^\xi - \sum_{j\neq i,\xi'} I_{ij}^{\xi\xi'}\mathcal O(\Omega_j)$, which is not dependent on $\Omega_i$. We can cast it into the coherent state representation as
\begin{align}
    \mathcal H_{{\rm loc},i}  &= \sum_{\al\beta} h_{\al\beta}(i) c_\al^*(\Omega_i) c_\beta(\Omega_i)
    =\sum_{\gm} \Lambda_{\gm}(i) |d_\gm(\Omega_i)|^2,
\end{align}
where the diagonalization is performed in the right-most side by the unitary matrix $V$: $d_\gm = \sum_\al V^\dg_{\gm\al}c_\al$.
It is apparent at this point that the energy does not change by the phase transformation $d_\gm \to d_\gm \epn^{\imu\theta_\gm}$ for any $\theta_\gm$, with which the coherent state is transformed as $\Omega_i \to \Omega_i'$.
The parameter $\theta_\gm$ is determined to minimize the norm of the inner product $\la \Omega_i|\Omega_i' \ra$ (see Ref.~\cite{supple} for more details).
This update makes it efficient to sample different configurations.
We note that the above procedure involving coherent state reproduces the over-relaxation update usually used for the SU(2) case.

\begin{figure}
    \centering
    \includegraphics[width=85mm]{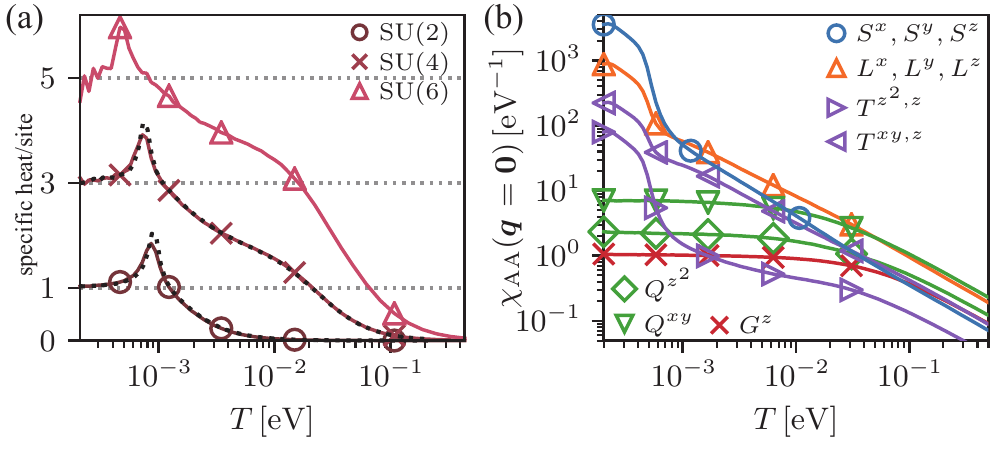}
        \caption{
        (a) Temperature dependence of the specific heat per site for the SU(2,4,6) models indicated by circles, crosses and triangles, respectively, which are obtained by classical Monte Carlo method.
        The black dotted lines for the SU(2,4) models are the results with $N_{\rm site}=256$ ($= 4 \times 4^3$).
        (b) Temperature dependence of the $\bm{q} = \bm{0}$ component of the diagonal susceptibilities at A-site for the SU(6) model.
    }
    \label{fig:n1_cl_tdep}
\end{figure}

We show the numerical result of the classical Monte Carlo in Fig.~\ref{fig:n1_cl_tdep}.
The calculation is performed for a finite-sized lattice with $N_{\rm site} = 108$ ($= 4\times 3^3$) atoms where the lattice is created using primitive translation vectors.
The temperature dependence of the specific heat is shown in Fig.~\ref{fig:n1_cl_tdep}(a) for the SU(2,4,6) models.
At low temperatures, the specific heat takes $2(N-1)\times \frac 1 2$ for the SU($N$) model, which satisfies the equipartition theorem.
Compared to the corresponding results of the mean-field calculation Fig.~\ref{fig:n1_mf_tdep}(d), every model has the suppressed transition temperatures down to $T_\mathrm{c} \sim 10^{-3}$~eV because of the incorporation of spatial fluctuations.
Figure~\ref{fig:n1_cl_tdep}(b) shows the $\bm{q} = \bm{0}$ component of the static susceptibilities for the SU(6) model.
For $T \lesssim 10^{-3}$~eV, each component of the magnetic moments takes the huge values, showing the feature of second-order phase transition.
The electric ($Q,G$) susceptibilities are characteristic for the SU(6) model and is absent in SU(2) cases.

\textit{Classical equation of motion}.---
Using the path-integral approach, our framework can further address the thermodynamic non-equilibrium state.
The equation of motion itself has already been derived by Zhang-Batista~\cite{Zhang2021}.
The derivation is based on the Heisenberg equation of motion of $\mathscr O_i^\xi$ which gives $N^2-1$ equations.
In terms of the parameters of the coherent states, on the other hand, we only need $2(N-1)$ equations.
Hence some of those equations should be redundant.
Here, we derive the $2(N-1)$ equations based on the principle of the least action of Eq.~\eqref{eq:action}~\cite{AuerbachBook1994,NagaosaBook1999}.
The resultant equation of motion for the local variable is given by
\begin{align}
\sum_q \mathcal B_{pq}(i) \frac{\partial \Omega_{qi}}{\partial\tau} 
= - \frac{\partial \mathcal H}{\partial \Omega_{pi}} ,
\label{eq:eom_path_integral}
\end{align}
where $\mathcal H = \la \bm \Omega| \mathscr H_{\rm eff} |\bm \Omega \ra$ and the Berry curvature matrix is defined by
\begin{align}
    \mathcal B_{pq}(i) &= \sum_\al \left(
    \frac{\partial c_\al^*(\Omega_i)}{\partial\Omega_{pi}}
    \frac{\partial c_\al  (\Omega_i)}{\partial\Omega_{qi}}
-
    \frac{\partial c_\al^*(\Omega_i)}{\partial\Omega_{qi}}
    \frac{\partial c_\al  (\Omega_i)}{\partial\Omega_{pi}}
\right),
\end{align}
with $p,q=1,\cdots, 2(N-1)$.
Changing the time variable as $\tau \to \imu t$, we obtain the real-time equation of motion.
Since the analytic form of the Berry curvature matrix is obtained once the specific coherent state is given in Eq.~\eqref{eq:coherent_state_def}, the even-dimension antisymmetric matrix $\mathcal B$ in Eq.~\eqref{eq:eom_path_integral} is 
easily inverted numerically.
Thus the explicit equation of motion is obtained for the $2(N-1)$ classical variables, and will be used for a non-equilibrium dynamics in a realistic setup.
The relation to the equations in Ref.~\cite{Zhang2021} is not apparent but can be deduced from the equation
\begin{align}
-\frac{\partial \mathcal O_i^\xi}{\partial \tau} = \sum_{p q} \mathcal B^{-1}_{pq}(i) \frac{\partial \mathcal O_i^\xi}{\partial \Omega_{pi}} 
\frac{\partial \mathcal H}{\partial \Omega_{qi}},
\end{align}
which derives from Eq.~\eqref{eq:eom_path_integral}.
The right-hand side is reminiscent of the commutator $[\mathscr O_i^\xi , \mathscr H_{\rm eff}]$.

\textit{Summary and outlook}.---
We have proposed the numerical calculation method for generic spin-orbital Mott insulators, and applied it to $5d$-pyrochlore oxides as a demonstration.
A detailed comparison between simulation results and experiments will provide us a deeper understanding of the Mott insulators, which leads to a design of functional materials.

\section*{Acknowledgement}

The authors thank R. Pohle for fruitful discussions.
This work was supported by KAKENHI Grants 
No.~19H01842, No.~21K03459 and No.~JP22J10620.

\bibliographystyle{apsrev4-2}
\bibliography{main.bbl}

\clearpage

\appendix

\makeatletter
\renewcommand{\thepage}{S\arabic{page}}
\renewcommand{\theequation}{S\arabic{equation}}
\renewcommand{\thefigure}{S\arabic{figure}}
\renewcommand{\thetable}{S\arabic{table}}
\makeatother

\setcounter{page}{1}
\setcounter{equation}{0}
\setcounter{table}{0}
\setcounter{figure}{0}

\noindent
{\bf SUPPLEMENTARY MATERIAL FOR \\
``Material-based analysis of spin-orbital Mott insulators''}
\\[2mm]
R. Iwazaki, H. Shinaoka, and S. Hoshino
\\[2mm]
(Dated: \today)

\section*{SM 1: Details of the first-principles calculation}
For constructing the tight-binding Hamiltonian, we used \texttt{Quantum ESPRESSO}~\cite{Giannozzi_2009,Giannozzi_2017} and \texttt{wannier90}~\cite{Wannier90}.
In the band calculations using \texttt{Quantum ESPRESSO}, we used pseudopotentials from \texttt{pslibrary} 1.0.0~\cite{DALCORSO2014337} and a kinetic energy cutoff of 75 Ry for the PAW method~\cite{PhysRevB.50.17953}.
The band calculations were done with the experimental lattice structure at 180 K: $a=10.1598$ \AA~and $x(\mathrm{O}_1)=0.319$~\cite{PhysRevB.63.195104}.
We constructed maximally localized Wannier functions using \texttt{wannier90} for the $t_\mathrm{2g}$ manifold.

\section*{SM 2: Effective Hamiltonian}
We write the Hamiltonian as $H=H_0+V$ where $V$ is treated by the perturbation theory.
The Hermitian effective Hamiltonian is given by~\cite{desCloizeaux1960,Klein1974,Durand1983}
\begin{align}
    \mathscr H_{\rm eff} &= (\Omega^\dg\Omega)^{-1/2} \Omega^\dg H \Omega (\Omega^\dg\Omega)^{-1/2},
\end{align}
where $\Omega$ is the wave operator determined by the operator equation
\begin{align}
    [H_0,\Omega] = - V\Omega + \Omega V \Omega.
\end{align}
We introduce the projection operators $P_0$ onto the model Hilbert space and also $Q_0=1-P_0$, which commute with $H_0$.
There are the relations~\cite{Durand1983}
\begin{align}
& P_0\Omega = P_0 , \ \ \  \Omega P_0 = \Omega.
\end{align}
Now we consider the perturbative expansion.
The square root is expanded as~\cite{Klein1974}
\begin{align}
    (\Omega^\dg\Omega)^{-1/2} &= P_0 + \sum_{j=1}^\infty \frac{(-1)^j}{2^{2j}} 
    \, 
    \comb{2j}{j} \, (\Omega^\dg \Omega - P_0)^j
\\
&\simeq
    P_0 - \frac 1 2 \Omega_1^\dg \Omega_1,
\end{align}
where $\Omega$ is expanded as $\Omega=\Omega_0 + \Omega_1 + \cdots$ and only the contributions up to second-order are kept.
The second-order effective Hamiltonian is given by
\begin{align}
    \mathscr H_{\rm eff}&=
    P_0 (H_0 + V) P_0
    +\frac{1}{2}( P_0 V\Omega_1 + \Omega_1^\dg V P_0 ).
\end{align}
The matrix element is evaluated as
\begin{align}
    &\la a | \mathscr H_{\rm eff} |b\ra  =
    \la a | (H_0 + V) |b\ra 
    \nonumber \\
    &\hspace{5mm}
    + \frac 1 2 \la a | V
    \Big(Q_0
    \frac{1}{E_a - H_0} + \frac{1}{E_b - H_0} Q_0
    \Big)V | b \ra,
\end{align}
where $|a,b\ra$ belong to the model Hilbert space.

\section*{SM 3: SU($N$) generators}
When we expand the effective Hamiltonian, we take the matrix representation $O_{\al\beta}^\xi$ as SU($N$) generators, where $\al,\beta,\xi \in \mathbb N$, $\al,\beta \in [1,N]$, and $\xi \in [0, N^2-1]$.
Just for convenience, we impose the Hermiticity, completeness, and orthonormality for the matrix basis:
\begin{align}
    &(\hat O^\xi )^\dg = \hat O^\xi,
    \\
    & \sum_\xi (O^\xi_{\al\beta})^* O^\xi_{\al'\beta'} = \delta_{\al\al'} \delta_{\beta\beta'},
    \\
    & \sum_{\al\beta} (O^\xi_{\al\beta})^* O^{\xi'}_{\al\beta} =
    \Tr \hat O^\xi \hat O^{\xi'}
     = \delta_{\xi\xi'},
\end{align}
where the hat ($\hat\ $) symbol represents a matrix with respect to the index $\al$.
Then the coupling constant satisfies $I_{ij}^{\xi\xi'}=I_{ji}^{\xi'\xi}\in \mathbb R$.
The matrix representation of SU($N$) generators consists of $N$ diagonal matrices and $N^2-N$ off-diagonal ones~\cite{GeorgiBook2000}.
The diagonal components are explicitly written as
\begin{align}
    &O_{\al\beta}^{\xi = 0}
    =
    \frac{1}{\sqrt{N}} \delta_{\al\beta},
    \\
    &O_{\al\beta}^\eta
    =
    \frac{1}{2\sqrt{\eta(\eta+1)}} \qty(
        \sum_{\zeta=1}^\eta \delta_{\al\zeta}\delta_{\beta\zeta}
        -
        \eta \delta_{\al, \eta+1} \delta_{\beta, \eta+1}
    ),
\end{align}
where $\eta \in [1, N-1]$.
$\hat O^0$ is proportional to the identity matrix.
The concrete forms of off-diagonal matrices are constructed by putting $1/\sqrt{2}$ or $-\imu/\sqrt{2}$ at one element of the upper triangular block.
The lower triangular block are determined from Hermiticity.

\section*{SM 4: Definition of the physical quantities}
In this section, we express the spin, magnetic orbital and quadrupole operators, whose expectation values are calculated in the main text.
The spin operator is defined by
\begin{align}
    \bm{S}_i
    =
    \frac{1}{2} \sum_{\gm\sg\sg'} c_{i\sg\gm}^\dg \bm{\sg}_{\sg\sg'} c_{i\sg'\gm},
\end{align}
where $\bm{\sg}$ is the Pauli matrix.
As for the orbital dependent quantities, we take the orbital basis $(|xy\ra, |yz\ra, |zx\ra)$ with the local coordinate of the $t_{2g}$ electrons where $z$-direction is along the local three-fold rotationally symmetric axis at each site.
The magnetic orbital is written as
\begin{align}
    \bm{L}_i
    =
    \sum_{\gm\gm'\sg} c_{i\sg\gm}^\dg \bm{\ell}_{\gm\gm'} c_{i\sg\gm'},
\end{align}
where
\begin{align}
    \hat{\ell}^x
    =
    \begin{pmatrix}
        0 & 0 & -\imu\\
        0 & 0 & 0\\
        \imu & 0 & 0
    \end{pmatrix},
    \hat{\ell}^y
    =
    \begin{pmatrix}
        0 & \imu & 0\\
        -\imu & 0 & 0\\
        0 & 0 & 0
    \end{pmatrix},
    \hat{\ell}^z
    =
    \begin{pmatrix}
        0 & 0 & 0\\
        0 & 0 & \imu\\
        0 & -\imu & 0
    \end{pmatrix}.
\end{align}
We have omitted the site index $i$ to make the notation simple.
We also write the quadrupole moment as
\begin{align}
    Q_i^\eta
    =
    \sum_{\gm\gm'\sg} c_{i\sg\gm}^\dg q_{\gm\gm'}^\eta c_{i\sg\gm'},
\end{align}
where the matrix representation $\hat{q}^\eta$ is constructed from $\hat{\ell}^\mu$ as
\begin{align}
    &\hat{q}^{x^2-y^2}
    =
    (\hat{\ell}^x)^2 - (\hat{\ell}^y)^2,
    \\
    &\hat{q}^{z^2}
    =
    \frac{1}{\sqrt{3}} (2(\hat{\ell}^z)^2 - (\hat{\ell}^x)^2 - (\hat{\ell}^y)^2),
    \\
    &\hat{q}^{xy}
    =
    \hat{\ell}^x \hat{\ell}^y + \hat{\ell}^y \hat{\ell}^x,
    \\
    &\hat{q}^{yz}
    =
    \hat{\ell}^y \hat{\ell}^z + \hat{\ell}^z \hat{\ell}^y,
    \\
    &\hat{q}^{zx}
    =
    \hat{\ell}^z \hat{\ell}^x + \hat{\ell}^x \hat{\ell}^z.
\end{align}
Combining the above matrices, we define another multipoles.
The electric dipole moment is written as
\begin{align}
    G_i^\mu =
    \frac 1 2 \sum_{\nu\lam} \ep_{\mu\nu\lam} \sum_{\gm\gm'\sg\sg'} c_{i\gm\sg}^\dg \ell_{\gm\gm'}^\nu \sg_{\sg\sg'}^\lam c_{i\gm'\sg'},
\end{align}
where $\ep_{\mu\nu\lam}$ is the completely anti-symmetric tensor.
At last, the magnetic octupole moment is written as
\begin{align}
    T_i^{\eta\mu} =
    \frac 1 2 \sum_{\gm\gm'\sg\sg'} c_{i\gm\sg}^\dg q_{\gm\gm'}^\eta \sg_{\sg\sg'}^\mu c_{i\gm\sg}.
\end{align}
The numerical results of the orbital dependent quantities $L, Q, G$ and $T$ shown in the main text are rotated to the globally defined axes (see the inset of Fig.~\ref{fig:ene_level} of the main text).

\section*{SM 5: Details of the classical model}

\subsection{Explicit form of the coherent state}
We consider the coherent state~\cite{Nemoto2000}
\begin{align}
    |\Omega \ra &= \sum_{\al=1}^N c_\al(\Omega) |\al\ra,
    \\
    c_\al(\Omega) &= \epn^{\imu \varphi_{\al-1}} \cos \xi_\al \prod_{\beta=1}^{\al-1} \sin \xi_\beta,
\end{align}
where $\varphi_{0,\cdots,N-1}$ and $\xi_{1,\cdots,N}$ with $\varphi_0 = 0$ and $\xi_N=0$.
Thus we have $2(N-1)$ parameters.
Here, we have omitted the site index $i$.
It follows that
\begin{align}
&\la \Omega |\Omega \ra = 1,
\\
&\frac{N!}{\pi^{N-1}}\int \diff \Omega\,
|\Omega\ra \la \Omega| = 1,
\\
& \diff \Omega = 
\prod_{\al=1}^{N-1}
\cos\xi_{\al} \sin^{2(N-\al)-1} \xi_{\al}
\ \diff \xi_\al  \diff \varphi_\al,
\\
&\braket{\Omega}{\partial\Omega} =
\imu \sum_{\al = 1}^N \abs{c_\al(\Omega)}^2 \partial\varphi_\al.
\label{eq:Berry_phase}
\end{align}
The parameters are $\xi_{1},\cdots, \xi_{N-1}\in [0,\pi/2]$ and $\varphi_{1},\cdots,\varphi_{N-1} \in [0,2\pi)$~\cite{Nemoto2000}.

\subsection{Over-relaxation update}
We provide a detailed explanation for the over-relaxation update.
We begin with the local Hamiltonian
\begin{align}
    \mathcal H_{{\rm loc}}  &= - \sum_\xi \tilde H^\xi \mathcal O(\Omega)
    \\
  &= \sum_{\al\beta} h_{\al\beta} c_\al^*(\Omega) c_\beta(\Omega)
    =\sum_{\gm} \Lambda_{\gm} |d_\gm(\Omega)|^2,
\end{align}
where the site index is omitted for simplicity.
The diagonalization is performed in the right-most side by the unitary matrix $V$: 
\begin{align}
    d_\gm (\Omega) = \sum_\al V^\dg_{\gm\al}c_\al (\Omega).
\end{align}
It is apparent at this point that the energy does not change by the phase transformation 
\begin{align}
d_\gm(\Omega) \to d_\gm \epn^{\imu\theta_\gm}
= \sum_\al \epn^{\imu\theta_\gm} V^\dg_{\gm \al} c_\al(\Omega)
\equiv \sum_\al V^\dg_{\gm \al} c_\al(\Omega'),
\end{align}
with which the coherent state is transformed as $\Omega \to \Omega'$.
For the efficient update to $\Omega'$, we would like to know the ``opposite side'' of $\Omega$ \cite{Creutz1987}.
For this purpose, we minimize the inner product defined by
\begin{align}
    \la \Omega | \Omega' \ra &= \sum_\gm \epn^{\imu\theta_\gm} |d_\gm(\Omega)|^2.
\end{align}
We define the norm
\begin{align}
    \mathcal N[\theta] &= |\la \Omega | \Omega' \ra|^2,
\end{align}
and we find that $\displaystyle \frac{\partial \mathcal N}{\partial \theta_\gm} = 0$ is satisfied if $\epn^{\imu\theta_\gm} = \pm 1 \equiv s_\gm$.
The set of signs ($s_1, \cdots , s_N$) is determined so as to minimize the norm $\mathcal N$.
We search for the solution by considering $2^N$ possibilities, which is same as the bipartitioning problem.
This procedure reproduces the over-relaxation update usually used for the SU(2) case.

\section*{
SM 6: Results for simple SU(2) model
}

Since the model with general $N$ is complicated, it is useful to summarize the results for the $N=2$ case as a benchmark.
Here we consider the single-orbital Hubbard model on the cubic lattice.
The Hamiltonian is given by
\begin{align}
\mathscr H &= -t \sum_{\la ij\ra \sg} c_{i\sg}^\dg c_{j\sg} + U\sum_i n_{i\ua} n_{i\da}.
\end{align}
Applying the second-order perturbation theory in the strong coupling limit, we obtain the effective Hamiltonian for $n=1$ as
\begin{align}
\mathscr H_{\rm eff} &= I \sum_{\la ij\ra} \sum_{\xi=1}^3 \mathscr O^\xi_i \mathscr O^\xi_j,
\end{align}
where $I=2t^2/U$ and
\begin{align}
\mathscr O^\xi = \sum_{\al,\beta=1}^2|\alpha\ra_i 
\, O^\xi_{\al\beta} \, {}_i\la \beta |.
\end{align}
The state vectors in model space are 
$|1\ra_i = c_{i\ua}^\dg |0\ra_i$ and 
$|2\ra_i = c_{i\da}^\dg |0\ra_i$, where $|0\ra_i$ is the vacuum at site $i$.
The matrices are constructed following the procedure in SM3:
\begin{align}
O^1 = \frac 1 {\sqrt 2}\begin{pmatrix}
    1 & 0\\
    0 & -1
\end{pmatrix}
,
O^2 = \frac 1 {\sqrt 2}\begin{pmatrix}
    0 & 1\\
    1 & 0
\end{pmatrix}
,
O^3 = \frac 1 {\sqrt 2}\begin{pmatrix}
    0 & -\imu \\
    \imu & 0
\end{pmatrix}.
\end{align}
We note that the above procedure is performed automatically in our framework.

    \begin{figure}
        \centering
        \includegraphics[width = 85mm]{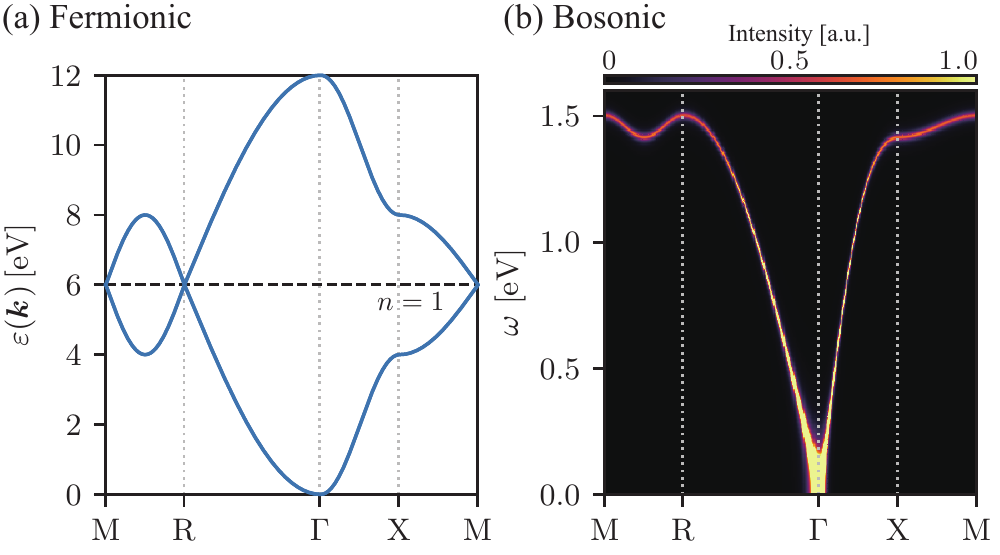}
        \caption{
        (a) Electronic band structure and (b) spin excitation spectrum ${\rm Im\,}\chi(\bm q,\omega)/\omega$ for the Heisenberg model on a cubic lattice.
        The parameters are chosen as 
            $t = 1$~eV, $U = 8$~eV and $T = 10^{-3}$~eV.
        }
        \label{fig:cub_band_qdep}
    \end{figure}

We take the two nearest neighbor atoms in the unit cell, which are labeled by the sublattice index A,B.
In this case, the antiferromagnetism occurs as a $\bm q=\bm 0$ state.
Here, we solve the Heisenberg model by the mean-field theory.
The spin excitation spectra are shown in Fig.~\ref{fig:cub_band_qdep}, which is an analog of Fig.~\ref{fig:n1_mf_qdep} of the main text. 
The gapless magnon mode around $\Gamma$ ($\bm q=\bm 0$) is clearly seen as expected.

The coherent state coefficients can also be explicitly written down ($\Omega_1 = \xi\in[0,\pi/2]$, $\Omega_2 = \varphi\in[0,2\pi)$):
\begin{align}
c_1(\Omega) &= \cos \xi,
\\
c_2(\Omega) &= \epn^{\imu \varphi} \sin \xi.
\end{align}
For the SU(2) case, $\varphi$ is interpreted as an azimuthal angle on the Bloch sphere, and $\theta = 2\xi \in [0,\pi]$ as a polar angle.
The Berry curvature matrix is given by
\begin{align}
    \mathcal B &= 
    \begin{pmatrix}
        0 & \imu \sin 2\xi \\
        -\imu \sin 2\xi & 0
    \end{pmatrix},
\end{align}
which leads to the Bloch's equation of motion for spin dynamics.

\vspace{10mm}
\noindent
{\bf \large References}
\\[1mm]
See the list of references in the main text.

\end{document}